\documentclass[]{iopart}
\usepackage{iopams}  
\expandafter\let\csname equation*\endcsname\relax
\expandafter\let\csname endequation*\endcsname\relax
\usepackage{amsmath,amssymb,amsfonts} 
\usepackage{graphicx} 
\usepackage{subfigure}
\newcommand{\be}{\begin{equation}} 
\newcommand{\ee}{\end{equation}}
\newcommand{\omhat}{\hat{\Omega}} 
\newcommand{\phat}{\hat{p}}
\newcommand{\hplus}{h_+}
 \newcommand{\hcross}{h_{\times}}
 
\newcommand{\lp}{\left(}
\newcommand{\rp}{\right)} 
\newcommand{\bb}{\begin{bmatrix}}
\newcommand{\eb}{\end{bmatrix}}

\begin{document}

\title{A Bayesian analysis pipeline for continuous GW sources in the PTA band}

\author{J.~A.~Ellis}

\address{Center for Gravitation, Cosmology and Astrophysics,
University of Wisconsin Milwaukee, Milwaukee WI, 53211}

\eads{justin.ellis18@gmail.com}

\begin{abstract} 
The direct detection of Gravitational Waves (GWs) by Pulsar Timing Arrays (PTAs) is very likely within the next decade. While the stochastic GW background is a promising candidate for detection it is also possible that single resolvable sources may be detectable as well. In this work we will focus on the detection and characterization of single GW sources from supermassive black hole binaries (SMBHBs). We introduce a fully Bayesian data analysis pipeline that is meant to carry out a search, characterization, and evaluation phase. This will allow us to rapidly locate the global maxima in parameter space, map out the posterior, and finally weigh the evidence of a GW detection through a Bayes Factor. Here we will make use of an adaptive metropolis (AM) algorithm and parallel tempering. We test this algorithm on realistic simulated data that are representative of modern PTAs.
\end{abstract}

\section{Introduction}

In the next few years pulsar timing arrays (PTAs) are expected
to detect gravitational waves (GWs) in the frequency range
$10^{-9}$ Hz--$10^{-7}$ Hz.  Potential sources of GWs in this frequency range include
supermassive black hole binary systems (SMBHBs) \cite{svc08}, 
cosmic (super)strings \cite{Olmez:2010bi}, inflation \cite{sa79}, and 
a first order phase transition at the QCD scale \cite{ccd+10}.
The community has thus far mostly focused on stochastic backgrounds produced 
by these sources, however; sufficiently nearby 
single SMBHBs may produce detectable continuous waves with periods on the order of 
years and masses in the range $10^8 M_{\odot}$--$10^9M_{\odot}$ \cite{wl03,svv09,sv10}. 
The concept of a PTA, an array of accurately timed 
millisecond pulsars, was first conceived of over two decades ago \cite{r89,fb90}. 
Twenty years later three main PTAs are in full operation: the 
North American Nanohertz Observatory for Gravitational waves (NANOGrav; \cite{jfl+09}), 
the Parkes Pulsar Timing Array (PPTA; \cite{m08}),
and the European Pulsar Timing Array (EPTA; \cite{jsk+08}).
The three PTAs collaborate  
to form the International Pulsar Timing Array (IPTA; \cite{haa+10}). 

A significant amount of work has gone into the detection problem for 
continuous GWs from SMBHBs. Both \cite{jll+04} and \cite{yhj+10}
use a Lomb-Scargle periodogram based approach to essentially measure
the excess power that a continuous GW would induce compared to a noise only
model. \cite{vl10} developed a Bayesian framework
aimed at the detection of GW memory in PTAs; however, the authors mention
that the methods presented could be used for continuous GW sources as well.
 Most recently, a maximized likelihood based approach has been
developed by \cite{bs12,esc12} and was later extended to include multiple 
resolvable sources in \cite{pbs+12}.

Many authors have focused on determining the parameter accuracy that we may
hope to extract from a future detection of a continuous GW from a SMBMB.
 \cite{sv10} use an Earth-term only signal model to perform a 
 study of SMBHB parameters that are measurable with PTAs using a Fisher matrix approach.
  \cite{cc10} have developed a Bayesian Markov Chain Monte-Carlo (MCMC) data analysis 
 algorithm for parameter estimation of a 
SMBHB system in which the pulsar term is taken into account 
in the detection scheme, thereby increasing the signal-to-noise-ratio (SNR) and improving the accuracy of
the GW source location on the sky. Recently, \cite{lwk+11} have developed parameter estimation 
techniques based on vector Ziv-Zakai bounds incorporating the pulsar 
term and have placed limits on the minimum detectable amplitude of a continuous 
GW source. In the aforementioned work, the authors also propose a method of combining timing parallax
measurements with single-source GW detections to improve pulsar distance measurements.

In this work we introduce a fully functional Bayesian pipeline aimed at both detection and parameter
estimation of single continuous GWs. To this end, we make use of MCMC augmented with Parallel
Tempering, an adaptive jump proposal scheme and thermodynamic integration for evidence evaluation.
 Previous work has made use the Fisher matrix
or similar techniques to either estimate parameter uncertainties or propose jumps in an MCMC algorithm.
Since it is known that the Fisher matrix is limited in use and only applies to large SNR
\cite{v08}, we choose to use an Adaptive Metropolis (AM) approach first developed in \cite{hst01,cj08} and
later applied to cosmology and GW parameter estimation in \cite{vrm+08,tg12,Taylor:2012vx}. 

The layout of the paper is as follows. In section \ref{sec:signal} we introduce the signal model and notation
used in this work. In section \ref{sec:implementation} we briefly review MCMC techniques, adaptive metropolis,
parallel tempering, thermodynamic integration and introduce our likelihood function and priors. In section \ref{sec:simData}
we introduce the semi-realistic simulated datasets that we use to test our algorithm. In section \ref{sec:mcmcTests} we test our 
algorithm on simulated data and make a few statements about the measurability of SMBMB parameters in realistic PTA data sets. 
Finally, we briefly mention future work and conclude in section \ref{sec:conc}.

\section{The signal model}
\label{sec:signal}

In general, pulsar timing residuals are defined as the difference of observed times-of-arrival (TOAs) of radio pulses and a deterministic timing model. In this section we will review the form of the residuals induced by 
a non-spinning SMBHB in a circular orbit and introduce our notation. The GW 
is a metric perturbation to flat space time defined in terms of its two polarizations as
\be
h_{ab}(t,\omhat)=e_{ab}^+(\omhat)\hplus(t,\omhat)+e_{ab}^{\times}(\omhat)\hcross(t,\omhat),
\ee
where $\omhat$ is the unit vector pointing from the GW source to the Solar
System Barycenter (SSB), $\hplus$, $\hcross$ and $e_{ab}^A$ ($A=+, \times$) are the  
polarization amplitudes and polarization tensors, respectively. The 
polarization tensors can be converted to the  
SSB by the following transformation. Following \cite{w87} we write
\begin{align}
e_{ab}^+(\omhat)&=\hat{m}_a\hat{m}_b-\hat{n}_a\hat{n}_b,\\
e_{ab}^{\times} (\omhat)&=\hat{m}_a\hat{n}_b+\hat{n}_a\hat{m}_b,
\end{align}
where
\begin{align}
\omhat &=-(\sin\theta\cos\varphi)\hat{x}-(\sin\theta\sin\varphi)\hat{y}-(\cos\theta)\hat{z},\\
\hat{m} &=-(\sin\varphi)\hat{x}+(\cos\varphi)\hat{y},\\
\hat{n} &=-(\cos\theta\cos\varphi)\hat{x}-(\cos\theta\sin\varphi)\hat{y}+(\sin\theta)\hat{z}.
\end{align}
In this coordinate system, $\theta=\pi/2-\delta$ and $\varphi=\alpha$ are the polar and azimuthal angles of the source, respectively, where $\delta$ 
and $\alpha$ are declination and right ascension in usual equatorial coordinates, where the North Celestial Pole is in the $\hat{z}$ direction and the Vernal Equinox is in the $\hat{x}$ direction.

We will write our GW induced pulsar timing residuals in the following form:
\be
s(t,\omhat)=F^{+}(\omhat)\Delta s_{+}(t)+F^{\times}(\omhat)\Delta s_{\times}(t),
\ee
where 
\be
\Delta s_{A}(t)=s_{A}(t_{p})-s_{A}(t_{e}),
\ee
and $t_{e}$ and $t_{p}$ are the times at which the GW passes the Earth\footnote{Technically, this is the time that the GW passes the SSB, however, following convention we will label this as the \emph{Earth} time and will later refer to the \emph{Earth}-term, keeping in mind that, in practice, all variables are referenced to the SSB.} and pulsar, respectively, and
the index $A\in\{+,\times\}$ labels polarizations. 
The functions $F^{A}(\omhat)$ are known as antenna pattern functions and are defined by
\begin{align}
F^{+}(\omhat)&=\frac{1}{2}\frac{(\hat{m}\cdot\phat)^{2}-(\hat{n}\cdot\phat)^{2}}{1+\omhat\cdot\phat}\\
F^{\times}(\omhat)&=\frac{(\hat{m}\cdot\phat)(\hat{n}\cdot\phat)}{1+\omhat\cdot\phat},
\end{align}
where $\phat$ is the unit vector pointing from the Earth to the pulsar. Also, from geometry we can write\footnote{Here we use geometrized units where $G=c=1$.}
\be
\label{eq:pTime}
t_{p}=t_{e}-L(1+\omhat\cdot\phat).
\ee
Given these definitions, we can write the GW contributions to the timing residuals as \cite{w87,cc10}
\begin{align}
\begin{split}
\label{eq:rplus}
s_{+}(t)&=\frac{\mathcal{M}^{5/3}}{D_L\omega(t)^{1/3}}\Big[-\sin[2\Phi(t)](1+\cos^{2}\iota)\cos2\psi\\
&-2\cos[2\Phi(t)]\cos\iota\sin2\psi\Big]
\end{split}\\
\begin{split}
\label{eq:rcross}
s_{\times}(t)&=\frac{\mathcal{M}^{5/3}}{D_L\omega(t)^{1/3}}\Big[-\sin[2\Phi(t)](1+\cos^{2}\iota)\sin2\psi\\
&+2\cos[2\Phi(t)]\cos\iota\cos2\psi\Big],
\end{split}
\end{align}
where $\psi$ is the GW polarization angle and $\iota$ is the inclination angle of the SMBHB. The orbital phase and 
frequency of the SMBHB are 
\be
\label{eq:phit}
\Phi(t)=\Phi_{0}+\frac{1}{32\mathcal{M}^{5/3}}\lp\omega_{0}^{-5/3}-\omega(t)^{-5/3}\rp
\ee
and
\be
\label{eq:worb}
\omega(t)=\omega_0\lp 1-\frac{256}{5}\mathcal{M}^{5/3}\omega_0^{8/3}t \rp^{-3/8}.
\ee
where $\Phi_{0}$ and $\omega_{0}$ are the initial values at the time of our first observation, the chirp mass is defined by $\mathcal{M}=(m_{1}m_{2})^{3/5}/(m_{1}+m_{2})^{1/5}$, where $m_{1}$ and $m_{2}$ are the masses of the two SMBHs, and $D_L$ is the luminosity distance to the source.  We can relate the GW frequency to the orbital frequency of the binary by $\omega_{\rm gw}=2\omega_0$ for circular orbits. Note that we use the observed redshifted values. For example, the chirp mass and orbital angular frequency in the rest frame are $\mathcal{M}_{r}=\mathcal{M}/(1+z)$ and $\omega_{r}=\omega_{0}(1+z)$, respectively, where $z$ is the cosmological redshift. 

Eqs. \ref{eq:phit} and \ref{eq:worb} are true in general  and can be applied when the frequency evolves appreciably over the total observing time. However, it is very useful to work under the assumption of slowly evolving binaries where $T_{\rm chirp}\gg T$, with $T$  the observing time and 
\be
T_{\rm chirp}=\frac{\omega_0}{\dot{\omega}}=3.2\times 10^5 \,{\rm yr}\lp \frac{\mathcal{M}}{10^8 \,{\rm M}_{\odot}} \rp^{-5/3} \lp \frac{f_0}{1\times 10^{-8} \,{\rm Hz}} \rp^{-8/3},
\ee
where 
\be
\dot{\omega}=\frac{96}{5}\mathcal{M}^{5/3}\omega_0^{11/3}.
\ee
Since typical PTA observations are on the order of 10--20 years and $T/T_{\rm chirp}\sim 10^{-4}$, this is a safe assumption for a broad range of masses and initial orbital frequencies of interest. With this approximation we can write the orbital frequency and phase for the earth term simply as
\begin{align}
\Phi_e(t)&=\Phi_0+\omega_0t\\
\omega_e(t)&=\omega_0.
\end{align}
However, for the pulsar term we are dealing with the retarded time of Eq. \ref{eq:pTime} and must include the first order corrections to the orbital frequency and phase
\begin{align}
\label{eq:Phipt}
\Phi_p(t)&=\Phi_0+\omega_0t-\omega_0 L(1+\omhat\cdot\phat)-\dot{\omega}L(1+\omhat\cdot\phat)t\\
\label{eq:omegap}
\omega_p(t)&=\omega_0-\dot{\omega}L(1+\omhat\cdot\phat),
\end{align}
where $L$ is on the order of a kpc and the last term in the pulsar phase containing $\dot{\omega}$ terms is responsible for any frequency evolution over the earth-pulsar light crossing time. As we will se later, writing the pulsar phase in this way will become very useful.

\section{Implementation}
\label{sec:implementation}

While Bayesian parameter estimation and model selection has been commonplace in  LIGO and LISA  \cite{cc05,vrm+08,vmr+09,lc09,l11,vma+12,aaa+13}, many PTA applications have been more frequentist in nature \cite{jll+04,jhl+05,abc+09,ych+11,yhj+10,bs12,esc12,pbs+12} and only recently has the Bayesian framework been put to use in the PTA context \cite{vlm+09,vl10,cc10,fl10,vhl12,esvh13,Lentati:2012xb,Taylor:2012vx}. Here we will briefly review Bayesian inference for clarity of notation. In the Bayesian framework, the data $d$ is assumed to be fixed and the parameters $\vec\theta$ that parameterize a hypothesis (or model) $\mathcal{H}$ are assumed to be randomly distributed. In this case, the data is used to update our prior knowledge of the hypothesis $p(\vec\theta,\mathcal{H})$ via Bayes theorem
\be
p(\vec\theta|d,\mathcal{H})=\frac{p(d|\vec\theta,\mathcal{H})p(\vec\theta,\mathcal{H})}{p(d|\mathcal{H})},
\ee
where $p(\vec\theta|d,\mathcal{H})$ is the posterior probability distribution, that is, the probability that the set of parameters $\vec\theta$ for hypothesis $\mathcal{H}$ could generate the given data $d$. In the above expression $p(d|\vec\theta,\mathcal{H})$ is the likelihood function, the probability that this dataset $d$ is drawn from a random distribution described by hypothesis $\mathcal{H}$ and parameterized by $\vec\theta$. Lastly, the prior $p(\vec\theta,\mathcal{H})$ encompasses any prior knowledge we have about the given hypothesis and $p(d|\mathcal{H})$ is the marginalized likelihood or evidence
\be
\label{eq:evidence}
p(d|\mathcal{H})=\int d\vec\theta\,p(d|\vec\theta,\mathcal{H})p(\vec\theta,\mathcal{H}).
\ee
For the purposes of parameter estimation we can safely ignore the evidence in Bayes theorem since it is just a normalizing factor that does not depend on the model parameters $\vec\theta$. However, if we want to perform model selection to claim a detection or compare different waveforms then the evidence is crucial. In this case we can make use of the Bayesian odds ratio between models ``$A$'' and ``$B$''
\be
\mathcal{O}=\frac{p(d|\mathcal{H}_A)}{p(d|\mathcal{H}_B)}\frac{p(\mathcal{H}_A)}{p(\mathcal{H}_B)},
\ee
where the first factor is known as the Bayes Factor which is our confidence in one model over the other based on the data (henceforth we will denote the Bayes factor as $\mathcal{B}$) and the second term is the prior odds ratio for models $A$ and $B$ which describes our prior belief in both models. In this paper we will only deal with the Bayes factor and assume that the prior odds ratio is unity.

In the following sections we will briefly review Markov Chain Monte Carlo (MCMC) and  describe some additional techniques used in our MCMC to help speed convergence and improve mixing. Finally, we will introduce our likelihood function and priors used in this work.

\subsection{Markov Chain Monte Carlo}

In this section we will quickly review the concept of MCMC. The appeal of MCMCs in general is that they sample directly from the posterior distribution and can efficiently explore the parameter space. The algorithm begins by specifying a point in some multidimensional parameter space $\vec{x}$. This point can be chosen at random from the prior or can be initialized in some other way if we have additional information about the posterior structure. From here, we propose a ``jump'' to a new point in parameter space, $\vec{y}$ via a jump proposal distribution function $q(\vec{y}|\vec{x})$. We then evaluate the posterior at this new point and accept the jump with probability $\alpha=\min(1,H)$ where $H$ is the Hastings ratio
\be
H_{\vec{x}\rightarrow\vec{y}}=\frac{p(\vec{y}|d)q(\vec{x}|\vec{y})}{p(\vec{x}|d)q(\vec{y}|\vec{x})}.
\ee
The Hastings ratio is constructed to ensure the reversibility in subsequent steps in the chain, a concept known as detailed balance. We repeat this process for many iterations until a convergence criteria is reached (i.e. autocorrelation time or Gelman Rubin R statistic \cite{gr92}) and the marginalized posterior pdfs of the parameters are simply the histograms of the parameter values in the chain. The choice of proposal distribution will be very important to achieve rapid convergence and we will explore this problem in the next section. 

\subsubsection{Adaptive Metrpolis}
\label{sec:adaptivemc}

Here we will outline an adaptive metropolis (AM) algorithm \cite{hst01}(hereafter HST01). As mentioned in the previous section the choice of jump proposal is very important for rapid convergence and adequate mixture of the chains. For this work we choose to make use of an adaptive scheme where the \emph{gaussian} proposal distribution is updated using the past history of the chain. By using the full past history of the chain this algorithm is indeed non Markovian but it is shown in HST01 that it retains the correct ergodic properties and thus will give unbiased samples from the posterior probability distribution. The algorithm is actually quite simple. First we use a multidimensional proposal distribution with diagonal covariance matrix  $C_0={\rm diag}(\epsilon\vec\lambda_{\rm start})$, where for this work we choose $\epsilon=10^{-6}$ and $\vec\lambda_{\rm start}$ is drawn from our prior distribution. By using this covariance matrix we assure that the initial jumps will be small (likely to be accepted) and thus we will begin to build up points for later adaptation. After some number of iterations, $\eta$, (for this work we choose $\eta=1000$) the covariance matrix at iteration $n$ becomes
\be
C_n=
\begin{cases}
C_0 & n\le\eta \\
s_d{\rm Cov}(\vec\lambda_0,\dots,\vec\lambda_{n-1}) & n>\eta\text{ and }{\rm mod}(n,\eta)=0,
\end{cases}
\ee
where $s_d$ is a parameter that depends on the dimension of the problem and ${\rm Cov}(\vec\lambda_0,\dots,\vec\lambda_{n-1})$ is the sample covariance matrix at the $n$th iteration of the algorithm. HST01 suggest a value of $s_d=2.4^2/n_{\rm dim}$, where $n_{\rm dim}$ is the dimension of the problem, however we have found that we need to use a smaller value to obtain optimal acceptance ratios around 25\% \cite{ggb+96}. As shown in the above equation, we do not perform the adaptation at every iteration of the chain but instead update the covariance matrix every $\eta$ iterations, which helps shorten the runtime of the algorithm. This adaptive method will help speed convergence as the jump proposal will begin to mimic the posterior and take into account any parameter correlations, however, in general, it will not help locate the global maximum of the posterior surface rapidly.

\subsubsection{Parallel tempering and thermodynamic integration}

A major problem with generic MCMC samplers is the tendency to get trapped in a local maxima. For a standard search it is unlikely that we will know a priori where the global maxima are located in parameter space, thus we must start our chain from a random point in the prior space. We want our algorithm to then quickly locate the global maxima in the parameter space. To accomplish this in a way that satisfies detailed balance we make use of parallel tempering. This technique involves different chains exploring the parameter space simultaneously, each with a different target distribution
\be
p(\vec\theta|d,\beta)=p(\vec\theta)p(d|\vec\theta)^{\beta},  
\ee 
where $\beta\le1$ is the inverse "temperature". This will essentially flatten out the likelihood surface allowing the chains to more freely explore the entire prior volume. The "hot" chains will inform the "colder" chains and vice versa by proposing parameter swaps between different temperatures. A parameter swap between the $i$th and $j$th temperature is accepted with probability $\alpha=\min(1,H)$, where the multi-temperature Hastings ratio is
\be
H_{i\rightarrow j}=\frac{p(d|\vec\theta_i,\beta_j)p(d|\vec\theta_j,\beta_i)}{p(d|\vec\theta_i,\beta_i)p(d|\vec\theta_j,\beta_j)}.
\ee
By swapping parameter states between different temperatures this ensures rapid location of the global maxima. The true posterior samples will come from the $\beta=1$ chain but the higher temperature chains can be used to evaluate the evidence via thermodynamic integration (see e.g. \cite{lc09} and references therein). Consider the evidence for a chain with temperature $1/\beta$ as part of a partition function
\be
\begin{split}
Z(\beta)&=\int d\vec\theta\, p(d|\vec\theta\mathcal{H},\beta)p(\vec\theta|\mathcal{H})\\
&=\int d\vec\theta\, p(d|\vec\theta,\mathcal{H})^{\beta}p(\vec\theta|\mathcal{H}).
\end{split}
\ee
Since the prior is independent of $\beta$, we can take the log and integrate over $\beta$ to obtain
\be
\label{eq:thermoInt}
\ln\,p(d|\mathcal{H})=\int_0^1d\beta\, \langle \ln\, p(d|\vec{\theta},\mathcal{H}) \rangle_{\beta},
\ee
where $\langle \ln\, p(d|\vec{\theta},\mathcal{H}) \rangle_{\beta}$ is the expectation value of the likelihood for the chain with temperature $1/\beta$. The expectation values are calculated over the post burn-in chains. In practice, it is important to choose a temperature ladder such that we explore the entire likelihood surface and recover the full integrand of Eq. \ref{eq:thermoInt}. For example, if we expect (or have injected) a signal with a given SNR, then the highest temperature chain will decrease the SNR by a factor $\sim 1/\sqrt{T}$, therefore; if we expect ${\rm SNR}=10$, then a maximum temperature of  $T_{\rm max}=100$ should be sufficient. However, we may need to use much higher temperatures to ensure the that above integral has converged \cite{lc09}. In this work our temperature ladder is constructed to be exponentially spaced with maximum temperature $T_{\rm max}={\rm SNR}^2$.

\subsubsection{Jump Proposals}

As mentioned in section \ref{sec:adaptivemc} we use an AM scheme to update the covariance matrix for multidimensional gaussian jumps. However since our parameter space is quite large ($8+N_{\rm psr}$) we do not always update all parameters simultaneously. In $\sim$70\% of jumps we will jump in subsets of correlated parameters such as the sky location parameters and pulsar distance as well as the chirp mass and distance. In $\sim$20\% of jumps we update all parameter simultaneously and in the remaining $\sim$10\% of jumps we choose one parameter at random and propose large jumps in parameter space.

In order to ensure proper mixing and exploration of our chains we have chosen to expand the parameter space in the following way. If we introduce the initial pulsar phase
\be
\phi_p=\omega_0L(1+\omhat\cdot\phat)
\ee 
and then solve for the pulsar distance
\be
L=\frac{\phi_p}{\omega_0(1+\omhat\cdot\phat)}+\frac{2\pi n}{\omega_0(1+\omhat\cdot\phat)}=L^{\rm small}+L^{\rm big},
\ee
where $n$ is the number of times the phase has wrapped around $2\pi$ (typically 1000s). By writing the distance to the pulsar in this fashion we can separate out the very small scale fluctuations ($L^{\rm small}$) that are important for coherence and are typically less than a pc, and the large scale fluctuations ($L^{\rm big}$) that are on the order of a kpc are important for determining the frequency evolution of the binary. These two components are essentially independent and explain physics on vastly different scales.  So now re-writing Eq. \ref{eq:Phipt} we have
\be
\Phi_p(t)=\Phi_0+\omega_0t-\phi_p-\dot{\omega}L(1+\omhat\cdot\phat)t,
\ee
where we jump in \emph{both} $\phi_p$ and $L\approx L^{\rm big}$.

\subsection{Likelihood and Priors}

Following \cite{vhl12,esvh13} we write the pulsar timing residuals in the linear approximation as
\be
\delta t=M\delta\boldsymbol{\xi}+n+s,
\ee
where $\delta t$ are the timing residuals, $M$ is the design matrix, $\delta\boldsymbol{\xi}$ is the parameter offset between the true pulsar timing parameters and our best fit parameters, $n$ is the noise present in the TOAs (radiometer noise, red noise, etc.), and $s$ is our continuous GW signal. Since $n$ is assumed to be gaussian we can write the likelihood function for a single pulsar as
\be
p(\delta t|\delta\boldsymbol{\xi},\vec\theta,\vec\lambda)=\frac{\exp\left[- \frac{1}{2} \lp \delta t -s -M\delta\boldsymbol{\xi} \rp^T C^{-1}\lp \delta t -s -M\delta\boldsymbol{\xi} \rp\right]}{\sqrt{(2\pi)^n\det C}},
\ee
where $\vec\theta$ are parameters that describe the noise in the pulsar residuals and $\vec\lambda$ the parameters that characterize the continuous GW signal. It was shown in \cite{vhl12} that this likelihood function can be marginalized over the timing model parameters $\delta\boldsymbol{\xi}$ to obtain
\be
p(\delta t|\vec\theta,\vec\lambda)=\frac{\exp\left[- \frac{1}{2} \lp \delta t -s \rp^T G(G^T CG)^{-1} G^T\lp \delta t -s\rp\right]}{\sqrt{(2\pi)^{n-m}\det(G^TCG)}},
\ee
where $G$ is an $n\times (n-m)$ matrix with $n$ the number of TOAs and $m$ the number of fitted parameters in the timing model. The derivation of $G$ can be found in \cite{vhl12} and will not be explored here. We can think of the matrix $G^T$ as a projection operator that projects our data $\delta t$ onto the null space of $M$, that is, it projects the data into a subspace orthogonal to the timing model fit. In this way we have fully taken into account the timing model fitting procedure.

For this work we will assume that the noise parameters $\vec\theta$ are know from some noise estimation done beforehand (see e.g. \cite{vhl12,esd+13}) and will only focus on characterizing the continuous GW parameters $\vec\lambda$. We will also assume that the residuals between pulsars are uncorrelated. In other words, we are assuming that the stochastic GW background will be negligible compared to the intrinsic noise in each pulsar. In general this is not likely to be a good assumption when we would expect a detection of a single GW source. The effects of omitting the correlations in the likelihood function are unknown and will be the subject of future work. Under these assumptions, the likelihood function for the full PTA can be written as
\be
p(\delta\mathbf{t}|\vec\lambda)=\prod_{\alpha=1}^{N_{\rm psr}}p(\delta t_{\alpha}|\vec\lambda_{\alpha}),
\ee
where $\delta t_{\alpha}$ and $\vec\lambda_{\alpha}$ and the residuals and model parameters for the $\alpha$th pulsar, respectively.
Since we are assuming the noise is fixed (and known) then we can write the log-likelihood ratio of a model with a single continuous GW to a model with just noise as
\be
\label{eq:lnlike}
\ln\,\Lambda=\sum_{\alpha}^{N_{\rm psr}}\left[ \lp \delta t_{\alpha}|s(\vec\lambda_{\alpha})\rp-\frac{1}{2}\lp s(\vec\lambda_{\alpha})|s(\vec\lambda_{\alpha})\rp\right],
\ee
where the inner product between two time-series $x$ and $y$ is
\be
(x|y)=x^TG(G^T CG)^{-1} G^Ty.
\ee

We choose flat priors on all angular parameters and flat priors in the log of the chirp mass, luminosity distance, and frequency of the GW. For the pulsar distance prior we use the current electromagnetic (EM) measurements either from timing parallax or Very Long Baseline Interferometry (VLBI) to contain the prior space as follows
\be
p(\vec L)=\prod_{\alpha=1}^{N_{\rm psr}}\frac{1}{\sqrt{2\pi \sigma_\alpha^2}}\exp\lp -\frac{(L_{\alpha}-L^{\rm EM}_{\alpha})^2}{2\sigma_{\alpha}^2} \rp,
\ee
where $L^{\rm EM}_{\alpha}$ is the best measured distance for the $\alpha$th pulsar  and $\sigma_{\alpha}$ is the 1-sigma uncertainty on that distance measurement.

\section{Simulated data sets}
\label{sec:simData}

In this work we will simulate ``toy model'' datasets that represent realistic yet optimistic present day residuals. We have chosen an array of 10 pulsars that are meant to represent the best 10 IPTA pulsars in terms of timing precision. The datasets have uneven sampling, varying error bars, and time spans corresponding to the real pulsar observing span. The data is summarized in Table \ref{tab:pulsars}.
\begin{table}[!h]
\caption{\label{tab:pulsars}Simulated IPTA pulsar datasets.  The RMS values are measured from the data with no injected signal. The pulsar distances are taken from \cite{vwc+12} if available. Otherwise the pulsar distances were taken from the ATNF catalog.}
\begin{indented}
\item[]\begin{tabular}{lccc}
\br
Pulsar Name & RMS [ns] & Time Span [yr] & Pulsar Distance [kpc]\\
\mr
J0437--4715 & 69 & 14.8 & $0.156\pm0.001$\\
J1909--3744 & 100 & 9.0 & $1.26\pm0.03$ \\
J1713+0747 & 136 & 18.3 &$1.05\pm0.06$\\
J1939+2134 & 141 & 16.3 & $5.0\pm2.0$\\
J1744--1134 & 366 & 16.9 & $0.42\pm0.02$\\
J1857+0943 & 402 & 14.9 & $0.9\pm0.2$\\
J1640+2224 & 410 & 14.9 & $1.19\pm0.24$\\
J2317+1439 & 412 & 14.9 & $1.89\pm0.38$\\
J1824-2452 & 602 & 5.7 & $3.6\pm0.72$\\
J0030+0451 & 792 & 12.7 & $0.28\pm0.1$\\
\br
\end{tabular}
\end{indented}
\end{table}
To create this data we use the mean RMS from the IPTA pulsars and draw each residual from a gaussian distribution centered on the RMS with a standard deviation of 50\% of the RMS. This way we are taking into account varying error bars and assuring that we only have gaussian white noise. We then simulate a continuous GW signal as in section \ref{sec:signal} and add it to our simulated noise. Finally, in an attempt to take into account the most important part of the timing model, we fit out a 2nd order polynomial from the data. The pulsar distances and uncertainties used in this analysis are the best measured values taken from \cite{vwc+12} if available, otherwise, we use the values from the Australia National Telescope Facility (ATNF) pulsar catalog\footnote[1]{\texttt{http://www.atnf.csiro.au/people/pulsar/psrcat/}} and assume a 20\% uncertainty. The rough cadence is chosen to simulate bi-monthly sampling. In order to present an idealistic yet plausible representation of current IPTA data sets, we have chosen to not include any intrinsic red noise which would only act to decrease sensitivity at low frequencies, therefore; the results presented here are likely to be optimistic.

\section{MCMC simulations}
\label{sec:mcmcTests}

In this section we wish to test the efficacy of our algorithm by injecting continuous GW signals into our simulated datasets described above. Although our main goal is to test our algorithm, we also wish to add a certain level of realism to these simulations. For this reason we have used mock IPTA datasets and will focus any astrophysical statements mostly to low SNR sources (SNR $\sim$ 7) as this represents a realistic possibility in the next decade. We also include injections at higher SNR and mimic these injections in ideal datasets (10 pulsars timed for 10 years all with 100 ns RMS drawn from an isotropic distribution on the sky) which have been used in previous parameter estimation work for PTAs \cite{cc10,sv10,lwk+11}.

Recent work has shown that there may be potential single GW source ``hot spots" in the Virgo, Fornax and Coma clusters \cite{sls+13}. Since our purpose here is only to illustrate the efficacy of our algorithm, we have randomly chosen to inject GW sources at the sky location corresponding to the Fornax cluster with a chirp mass of $\mathcal{M}=7\times 10^8 M_{\odot}$ and initial orbital period of 3.16 yr. The distance to the GW source is then scaled such that we achieve the desired SNR defined by 
\be
{\rm SNR}^2=\sum_{\alpha}\lp s(\vec\lambda_{\rm inj})|s(\vec\lambda_{\rm inj})\rp_{\alpha},
\ee
 where the sum is over the number of pulsars and $\vec\lambda_{\rm inj}$ are the injected source parameters. This choice of injected parameters is justified since the amplitude of our GW induced residuals scales as $\mathcal{M}^{5/3}\omega^{-1/3}$ and the stochastic GWB and other potential red noise sources will lower our sensitivity at lower frequencies. Therefore, we are likely to detect a source with high chirp mass and high frequency. See table \ref{tab:gwPars} for a list of the different GW sources and parameters used in this work. 
 \begin{table}[!h]
\caption{\label{tab:gwPars} Simulated GW source parameters. These sources are injected at the sky location of the Fornax cluster and the distance is scaled such that we achieve the desired SNR.}
\begin{tabular}{@{}ccccccccc}
\br
SNR & $\theta$ [rad] & $\varphi$ [rad] & $\psi$ [rad] & $\iota$ [rad] & $\Phi_0$ [rad] & $\mathcal{M}$ [$M_{\odot}$] & $D_L$ [Mpc] & $f_{\rm gw}$ [Hz]\\
\mr
7 & 2.17 & 0.95 & 1.26 & 1.57 & 0.99 & $7.0\times 10^8$ & 223.4 & $2\times 10^{-8}$\\
14 & 2.17 & 0.95 & 1.26 & 1.57 & 0.99 & $7.0\times 10^8$ & 111.7 & $2\times 10^{-8}$\\
20 & 2.17 & 0.95 & 1.26 & 1.57 & 0.99 & $7.0\times 10^8$ & 78.2 & $2\times 10^{-8}$\\
\br
\end{tabular}
\end{table}
 For each source, the same noise realization was used so that relative parameter accuracies do not depend on this specific noise realization. In general we would like to do a much more detailed analysis with many different noise realizations and many different injected sources. Indeed, this will be the subject of future work, however; here we simply want to test the various steps of our algorithm, that is, the search phase where we find the global maxima in the multi-dimensional parameter space, the sampling phase where we obtain samples from the underlying posterior distribution, and finally the evaluation phase where we compute the evidence and Bayes factors to make choices about detection.
 
 \subsection{Searching for global maxima}
 
\begin{figure}[!h]
  \begin{center}
	\includegraphics[scale=1.0]{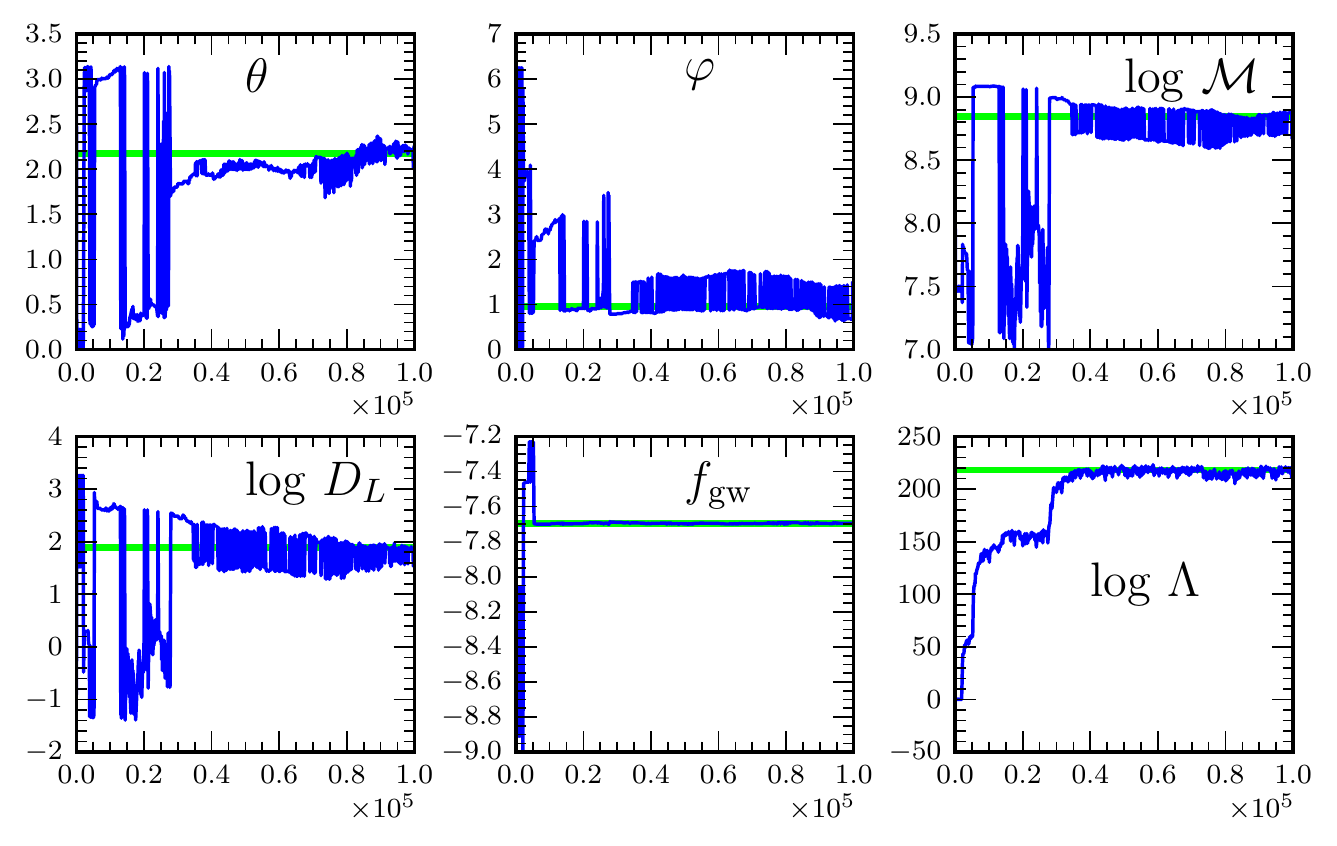}	
   \end{center}
  \caption{\label{fig:fornaxTrace} Trace plots for the measurable parameters (the inclination angle, initial phase and polarization angle are not well constrained for this realization) for an SNR=20 injection for the first $10^5$ steps. In all cases the black(green) line represents the injected parameters and the gray(blue) is the chain trace. We can see that the parallel tempering scheme has allowed us to locate the global maxima of the log-likelihood and all parameters within the first $\sim 6\times 10^4$ steps. (colour figures online.)}
\end{figure}

 Since we have little information about the SMBMB population, we want to carry out a blind search of the parameter space making no assumptions about the underlying SMBMB source parameters. Therefore, it is very important that our algorithm be able to quickly find the global maxima of the log-likelihood function and the true parameters so that the sampling process can begin. The trace plots of one ${\rm SNR}=20$ injection is shown in \Fref{fig:fornaxTrace}
where we have plotted the measurable parameters (excluding the pulsar distance) as well as the log-likelihood as a function of chain iteration for the $T=1$ chain. Here we do not plot the polarization angle, initial phase, or inclination angle as they are not well constrained by the data and contribute little to the overall log-likelihood for this case. We can see from the figure that the algorithm has correctly found the true source parameters within the first $\sim 6\times 10^4$ MCMC iterations. We note that the true value of the frequency is found quickly (within the first $10^4$ steps of the algorithm) and we reach the true value of the log-likelihood within the first $4\times 10^4$ steps. There are several ways that we could improve this step such as choosing a more suitable starting jump proposal distribution before starting adaptation or even starting adaptation sooner, however for the purpose of this work we believe that this is sufficient as the algorithm can still collect $\sim 2\times 10^6$ samples with 8 chains  in about 4 hours running on a 2.7 GHz quad core MacBook Pro. It is also important to note that in practice we will have carried out a simpler search algorithm such as an $\mathcal{F}$-statistic \cite{bs12,esc12,pbs+12} search prior to this Bayesian analysis. If any signal is detected, then we will have a very good idea of the frequency of the GW source and can therefore seed our MCMC algorithm much closer to the true value. Since the frequency contributes heavily to the log-likelihood, it is likely that this could reduce the number of samples required for this search phase by at least an order of magnitude.

 \subsection{Sampling and parameter estimation}
For each injected source we run 4 serial chains all with 8 temperatures and starting positions chosen at random from the prior, thereby assuring that our algorithm can indeed locate the global maxima. Each serial chain was run for $\sim 1.5\times 10^6$ iterations and 25\% of each chain was discarded as burn in. The resulting post burn-in chains were then concatenated to form a single chain with $\sim 4.5\times 10^6$ posterior samples. 

\begin{figure}[!h]
  \begin{center}
	\includegraphics[scale=1.0]{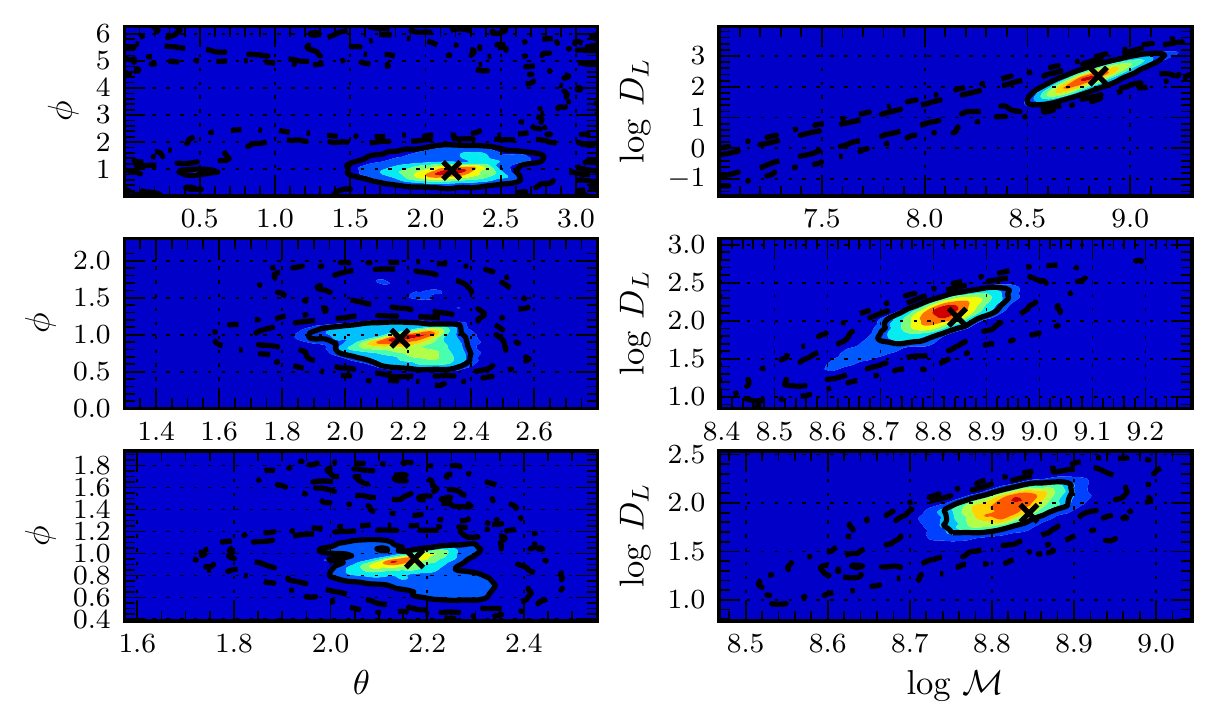}
   \end{center}
  \caption{\label{fig:fornaxPost} Marginalized 2-D posterior pdfs in the sky coordinates ($\theta,\phi$) and the log of the chirp mass and distance ($\log\,\mathcal{M},\log\,D_L$) for injected SNRs of 7, 14, and 20 shown from top to bottom. Here the injected GW source is in the direction of the Fornax cluster with chirp mass $\mathcal{M}=7\times 10^8 M_{\odot}$. The distance to the source is varied to achieve the desired SNR. Here the ``$\times$" marker indicates the injected parameters and the solid, dashed and dot-dashed lines represent the 1, 2, and 3 sigma credible regions, respectively. (colour figures online.)}
\end{figure}

\Fref{fig:fornaxPost} shows marginalized 2-D posterior pdfs of the sky coordinates ($\theta,\varphi$) and the log of the chirp mass and distance ($\log\,\mathcal{M},\log\,D_L$) for injected SNRs of 7, 14, and 20  (shown from top to bottom) for a source injected at the sky position of the Fornax cluster. The ``$\times$" marker indicates the injected parameters and the solid, dashed and dot-dashed lines represent the 1, 2, and 3 sigma credible regions, respectively. The first thing to note from this figure is that the injected value lies well within the 1-sigma credible regions in all three cases. We also note that since we have injected a relatively high mass and high frequency GW, we can measure $\dot{f}$ and therefore; we can can break the degeneracy between chirp mass and distance as is seen in the plots on the right in the above figure. Since we know the true injected values, it is possible to determine how much each pulsar contributes to the log-likelihood function. For the aforementioned injection, four pulsars contribute more than 1\% to the likelihood function for the SNR 7 injection and only three pulsars contribute more than 1\% to the likelihood function  for the SNR 14 and 20 injections. While this number does depend on the relative sky locations of the pulsars and the GW source as well as the specific noise realization, it is also a very strong function of the RMS of the noise in each pulsar ($\langle \ln\, \Lambda \rangle\propto \sigma_{\rm RMS}^{-2}$). In fact, we can see the results of this in \Fref{fig:fornaxPost} where there is a bit of multi modality in the posterior for sky position because we essentially only have three and four baselines (detectors) for the SNR 7 and 14 and 20 cases, respectively. 

This type of parameter degeneracy due to the small number of baselines differs from previous parameter estimation studies \cite{cc10,sv10,lwk+11} where the simulated PTA consisted of a large number (20 or more) of pulsars all timed to the same accuracy. For this reason, quoted SMBHB parameter accuracies that can be obtained from PTAs should be interpreted cautiously as it is extremely unlikely that future era PTAs will even approach this ideal situation. To illustrate this point we have also simulated an ideal data set of 10 pulsars drawn uniformly on the sky with 100 ns RMS in each with baselines of 10 years. We also chose distances drawn uniformly from the range $L\in [0.5,1,5]$ kpc with 10\% uncertainties. We have then used the same injection as in the simulated IPTA data at SNRs of 7, 14 and 20. The sky resolution \cite{c98} and fractional uncertainties on the chirp mass and distance for the simulated IPTA dataset are $\Delta \Omega=(2357.9,122.2,67.2)$ deg$^2$, $\Delta \mathcal{M}/\mathcal{M}=(48.8\%,9.5\%,6.3\%)$ and $\Delta D_L/D_L=(81.2\%,28.2\%,19.9\%)$, respectively. Whereas, for our ideal simulated datasets the corresponding values are $\Delta \Omega=(1085.9,23.7,12.8)$ deg$^2$, $\Delta \mathcal{M}/\mathcal{M}=(47.9\%,4.4\%,3.0\%)$ and $\Delta D_L/D_L=(79.7\%,15.9\%,13.2\%)$, respectively. Again, these results are not robust, in that we have only done one injection (with varying SNR) into one noise realization. Nonetheless, it should be clear that our simulated IPTA data do not yield nearly as precise sky resolution or chirp mass and distance fractional uncertainties as an ideal data set.

\subsection{Evaluating the evidence}

After we have carried out our parallel tempering MCMC search we can make use of the different temperature chains to calculate the evidence integral via Eq. \ref{eq:thermoInt}. Since we have measured the noise parameters before conducting our search, we use the log-likelihood ratio defined in Eq. \ref{eq:lnlike} as our log-likelihood. By doing this we can compute the Bayes factor comparing our GW and noise models simply by calculating the evidence using the log-likelihood ratio.
\begin{figure}[!h]
  \begin{center}
	\includegraphics[scale=1.0]{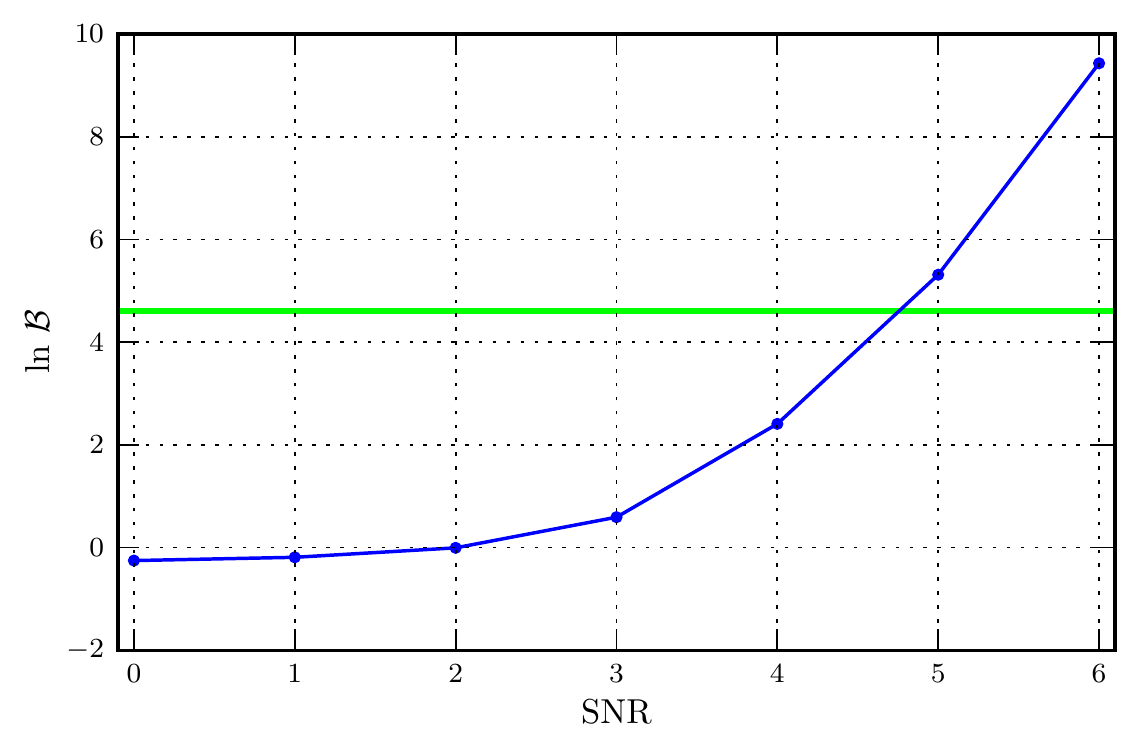}
   \end{center}
  \caption{\label{fig:fornaxBF} Log of the Bayes factor plotted against injected SNR for the same signal and noise realization. The gray(green) horizontal line is the threshold in the log of the Bayes factor in which we can claim a detection and the black(blue) points are the log Bayes factor calculated from  thermodynamic integration. (colour figures online.)}
\end{figure}
\Fref{fig:fornaxBF} shows the log of the Bayes factor computed from thermodynamic integration for injections at different SNRs. Here we have done injections into the same noise realization of out simulated IPTA data using the same GW source (again with the distance scaled to give the desired SNR) as above. The MCMC sampler was run with 10 temperature chains for $\sim 2\times 10^6$ iterations. In the figure, the gray(green) vertical line represents our threshold in the log of the Bayes factor of $\ln 100$, above which there is decisive evidence for a GW source \cite{j61} and the black(blue) points are the computed log Bayes factor for each injection. There are two important things to note. First,  notice that the log of the Bayes factor is above the threshold for injected sources with ${\rm SNR}\ge 5$ which agrees well with a frequentist interpretation of the SNR as a detection statistic in gaussian noise, where 5-sigma is usually required for a definitive detection. Secondly, as was discussed in \cite{lc09}, the Bayes factor is about unity for the zero to low SNR injections. This is because of the nature of the question that we are asking. In this case we are asking ``Is there evidence for \emph{any} continuous GW source in the data?". Framed in this way, the result makes perfect sense because a low SNR signal is nearly indistinguishable from pure noise, therefore the odds of a low SNR GW are about 50/50 indicated by a Bayes factor of 1. If we were to ask the question ``Is there a continuous GW source with SNR$\ge 5$ in the data?", then we would expect the Bayes factor to become much less than unity at low SNR.

\section{Conclusions and future work}
\label{sec:conc}

We have developed a robust MCMC algorithm that makes use of an Adaptive Metropolis scheme and parallel tempering for use in PTA detection and parameter estimation of single sources of GWs from SMBHBs. We have tested the algorithm on a fairly realistic simulated IPTA dataset that has many of the features of real data including uneven sampling, varying error bars and overall noise levels, poor pulsar distance measurement uncertainty and varying data span. For comparison we have also run the algorithm on ideal datasets, similar to those that have been considered in the literature. The algorithm has shown to perform well in the three stages of our Bayesian analysis pipeline, namely the search, sampling and evaluation phase. When seeded from a random point in parameter space, the algorithm can quickly locate the global maxima through the use of parallel tempering. Posterior samples are then collected efficiently through the use of Adaptive Metropolis and special jump proposals in an extended parameter space. Finally, we have shown that this algorithm can also be used for detection through the use of parallel tempering and thermodynamic integration to calculate the Bayesian evidence.

From the few simulations and comparisons of realistic vs. ideal data done in this work we can say that parameter estimation from current generation PTAs, counter to previous work on the subject, is likely to suffer due to the fact that few pulsars contribute to the total network SNR, resulting in a lower number of effective ``detectors" than the number of pulsars in the array. A much more detailed study of the parameter estimation problem in current generation PTAs with more realistic noise models (including effects such as time varying Dispersion Measure) is underway and will be the subject of a future paper.

In the future we hope to further modify this algorithm to incorporate intrinsic noise and stochastic background parameters in the search. In principle this is possible, however; this will result in a much larger parameter space and much more computationally demanding problem. To this end we plan on improving the algorithm to make it more efficient by implementing better initial jump proposals before the adaptation begins and also implementing an inter-chain adaptation scheme \cite{crr+09} which allows for a much more efficient parallelization than running many independent chains in serial. Finally, we also plan to allow for multiple single GW sources in our data.

\ack
We thank the members of the NANOGrav detection working  group for their comments and support. We would also like to thank Xavier Siemens, Richard O'Shaughnessy, Leslie Wade, and Jolien Creighton for many useful discussions. This work was partially funded through the Wisconsin Space Grant Consortium and the NSF through PIRE award number 0968126.

\section*{References}
\bibliographystyle{iopart-num} 
\bibliography{apjjabb,bib}

\end{document}